\documentclass[letter]{aa}

\usepackage{mathtools}

\usepackage{graphicx}
\usepackage{txfonts}

\begin{document}

   \title{Dust survival in harsh environments}

   \subtitle{Is photo-evaporation an important destruction mechanism?}

   \author{A. Nanni
          \inst{1,2}
          \and S. Cristallo\inst{2,3} 
          \and D. Donevski\inst{1,4,5}
          \and M. J. Michałowski \inst{6}
          \and M. Romano\inst{1,7}
          \and P. Sawant\inst{1}}

   \institute{National Centre for Nuclear Research, ul. Pasteura 7, 02-093 Warsaw, Poland
   \email{ambra.nanni@ncbj.gov.pl}
\and
INAF - Osservatorio astronomico d'Abruzzo, Via Maggini SNC, 64100, Teramo, Italy
\and
INFN - Sezione di Perugia, Via A. Pascoli SNC, 06123, Perugia, Italy
\and
SISSA, Via Bonomea 265, Trieste, Italy
\and
IFPU – Institute for fundamental physics of the Universe, Via Beirut 2, 34014 Trieste, Italy
\and
Astronomical Observatory Institute, Faculty of Physics, Adam Mickiewicz University, ul.~S{\l}oneczna 36, 60-286 Pozna{\'n}, Poland
\and 
INAF – Osservatorio Astronomico di Padova, Vicolo dell’Osservatorio 5, 35122 Padova, Italy
             }

   \date{Received ...; accepted...}

  \abstract
   {}
   {We investigate the role of photo-evaporation of dust exposed to the radiation field from hot young stars and planetary nebulae (PNe) as a possible destruction mechanism of dust grains in the interstellar medium (ISM).}
   {We estimate photo-evaporation induced by the feedback of individual or clustered young stars, of PNe and in the presence of a variable radiation field scaled with the interstellar radiation field. For PNe we investigate dust photo-evaporation of both dust grains already present in the ISM as well as those formed in the last phases of the evolution of thermally pulsing asymptotic giant branch (TP-AGB) stars. We include dust photo-evaporation rate in models of dust evolution in galaxies for different assumptions of the dust growth scenario, dust-to-gas ratios, star formation histories, and initial mass functions of the stars.}
   {For all the cases considered, we find that both photo-evaporation from young stars and from PNe are negligible with respect to other dust removal processes such as destruction from supernovae shocks, astration and possibly outflow. Grains are stable against photo-evaporation if they are exposed to a radiation field which is up to 10$^7$ times the interstellar radiation field.}
   {Dust grains of size $\geq 0.01$~$\mu m$ are not efficiently destroyed by photo-evaporation also in the presence of a strong radiation field.}


   \maketitle

\section{Introduction}
Formation and survival of dust grains in galaxies have paramount implications for many astrophysical processes. Dust grains are responsible for the build-up of different molecules in space, and for gas cooling, promoting the formation of stars (e.g., \citealt{Cuppen17}). Dust absorbs light mostly in the ultraviolet (UV) and visible wavelength and re-emits this energy in the infrared. Therefore, the physical quantities derived from the spectral energy distribution (SED) fitting, e.g. the star formation rate (SFR), must take the effect of dust into account.
Despite the significance of dust formation, destruction, and survival in the interstellar medium (ISM), such mechanisms remain controversial.

UV photons impinging on dust grains remove electrons from their surface and heat the gas. Through such a process grains become positively charged \citep{Weingartner01b}.
Photo-evaporation of dust grains is suggested to be relevant for the life-cycle of polycyclic aromatic hydrocarbons \citep[PAHs,][]{Demyk11}. Indeed, the ISM may be enriched of small molecules from the photo-evaporation of PAHs, while in photo-dissociation regions PAHs may be created when UV photons and/or shock waves break down carbon grains \citep{Cesarsky00, Berne07}. Nevertheless, the role of dust photo-evaporation as a possible efficient destruction mechanism of dust grains in the ISM has not been equally well explored so far.
Dust grains undergo photo-evaporation due to the intense radiation field in galaxies with high SFRs. In spite of that, in simulations of dust evolution in galaxies, the role of such a mechanism in relation to other processes, such as astration and supernovae (SN) shocks and large-scale galactic outflows has not been thoroughly assessed yet. In low-metallicity galaxies and the early Universe, where stars are typically hotter than at solar-like metallicity, and where the formation of massive stars with intense radiation fields may be favoured \citep{Marks12}, photo-evaporation may be potentially relevant.

Recent observational studies show that the specific mass of dust ($sM_{\rm dust}$=$M_{\rm dust}/M_{\star}$, where $M_{\rm dust}$ and $M_{\star}$ are the masses of dust and of stars, respectively) rises quickly at young ages and then decreases with age. Different classes of galaxies show evidence of this trend: massive and dwarf star-forming galaxies in the local Universe \citep{deVis19, Nanni20, Galliano21, Casasola22}, dusty galaxies and Lyman Break Galaxies identified at very high redshifts ($2<z<6$, \citealt{Burgarella20, Donevski20, Kokorev21, Burgarella22}), as well as quiescent but dusty galaxies at low and intermediate redshifts (\cite{Gobat19, Michalowski19, Donevski23, Lesniewska23}). The core of the correlation between the $sM_{\rm dust}$ and the specific star formation rate ($sSFR=SFR/M_\star$) is believed to be an age-evolutionary sequence.

From a theoretical viewpoint, various models tracking the dust, gas and metal content of galaxies that include recipes for dust formation, grain growth and destruction, as well as inflows and outflows, are able to reproduce the observed decline of $sM_{\rm dust}$ with $sSFR$ \citep{Li19, Pantoni19, deVis19, Nanni20, Galliano21, Burgarella22}. The majority of models find strong outflows driven by stellar feedback and SN explosions to explain the decline of $sM_{\rm dust}$ with age. Predicted outflow efficiencies ($ML=\dot{M}/SFR$, where $\dot{M}$ represents the mass outflow rate) are up to 80. In contrast to what simulations predict, recent observational works on both local and high-z sources (e.g. \citet{Ginolfi20, Romano23, Salak23}) found a typical value of $ML\approx1$. Using a novel spectral selection, \citet{Donevski23} examined quiescent galaxies that showed no signs of energetic feedback from embedded active galactic nuclei. A significant scatter follows an anti-correlation between $sM_{\rm dust}$ and age, suggesting distinct dust removal pathways over a range of timescales. Dust destruction from feedback from Type Ia SNe \citep{Li20}, planetary nebulae (PNe) \citep{Lesniewska23} or heating from winds of thermally pulsing asymptotic giant branch (TP-AGB) stars have also been proposed to explain the observed decline of $sM_{\rm dust}$ with age in quiescent galaxies \citep{Conroy15}.

This all motivates us to re-examine under what conditions dust can survive in galaxies and what processes are dominant in producing the observed anti-correlation of $sM_{\rm dust}$ and stellar age. Along with usually considered dust removal processes in the ISM such as SN shocks, astration, and outflow we additionally probe whether photo-evaporation due to the intense radiation input from massive stars and PNe can be an efficient destruction process for dust grains. 

\section{Method: dust evolution model}\label{method}
In this work, we investigate the efficiency of dust photo-evaporation by including this process in calculations that follow dust evolution in the ISM of galaxies. 
We here do not consider all the details of dust condensation and destruction during the process of star formation or PNe, but only the effect of radiation feedback from stars already in their main sequence, i.e. in HII regions, or in the PN phase.
\subsection{General framework}
In order to model gas and dust evolution in the ISM of galaxies, we first compute the chemical gas enrichment from stars by adopting the one-zone chemical evolution code \textsc{omega} \citep[][One-zone Model for the Evolution of GAlaxies]{Cote17, Ritter18} which includes population III  stars \citep{Heger10}, Type II \citep{LC18, Prantzos18} and Type Ia supernovae  \citep[SNe,][]{Iwamoto99} and TP-AGB stars \citep{Cristallo15}. We therefore compute the evolution of the gas mass $M_{\rm gas}$ as well as the the mass of each metal species, $M_{\rm gas, i}$, and the mass of the dust species, j, either silicates (olivine and pyroxene) or carbon, $M_{\rm dust, j}$, similarly to other works in the literature \citep[e.g.][]{Dwek98, Calura08, Asano13, Ginolfi18, deVis19, Nanni20, Galliano21}. 

If only inflow is neglected we obtain:
\begin{equation}\label{Mgas}
    \frac{dM_{\rm gas}}{dt}=\frac{dM^{\rm SP}_{\rm gas, ej}}{dt} -SFR -ML \times SFR - \sum_{\rm j} \frac{dM_{\rm dust, j}}{dt},
\end{equation}

where the last term takes into account the amount of metals locked into dust grains.

\begin{equation}\label{MZ}
\begin{multlined}
    \frac{dM_{\rm gas, i}}{dt}=\frac{dM^{\rm SP}_{\rm gas, ej, i}}{dt} -SFR \frac{M_{\rm gas,i}}{M_{\rm gas}} - ML\times SFR \frac{M_{\rm gas,i}}{M_{\rm gas}} - \\
    -\sum_{\rm j}n_{\rm i}\frac{m_{\rm i}}{m_{\rm dust, j}}\frac{dM_{\rm dust, j}}{dt},
\end{multlined}
\end{equation}

where the first two terms of each equation are computed by \textsc{omega} to which we refer for all the details. The first term represents the gas return from stellar population (SP). The initial mass function (IMF) of stars is assumed to be constant with time. The integral is performed between the minimum ($M_{\rm L}=0.8$~M$_\odot$) and the maximum ($M_{\rm U}=120$~M$_\odot$) mass of stars. The IMF is normalised in such a way that:
\begin{equation}\label{eq:imf_norm}
    \int^{M_{\rm U}}_{\rm M_{\rm L}} m IMF(m) dm = 1~M_\odot
\end{equation}
We here test two different IMFs: the Chabrier IMF \citep{Chabrier03} and a top-heavy IMF with the form:
\begin{equation}
    IMF(m) \propto m^{-\alpha},
\end{equation}
with $\alpha=1.35$. Such an IMF favours the formation of massive stars and can impact the efficiency of the destruction processes involved.

The second term of the Eqs.~\ref{Mgas} and \ref{MZ} is the
astration of gas and metals due to formation of stars while the last term is gas removal from the ISM operated by outflows. The last term of Eq.~\ref{MZ} represents metal depletion from the gas phase into dust grains, where $n_{\rm i}$ is the number of atoms of the element i in the monomer of the dust species j, and $m_{\rm i}$ and $m_{\rm dust, j}$ are the atomic mass of the element i and the dust monomer j, respectively.

The evolution of dust grains of species $j$ is computed as:
\begin{equation}\label{eq:dust}
\begin{multlined}
    \frac{dM_{\rm dust, j}}{dt}=\frac{dM^{\rm SP}_{\rm dust, ej, j}}{dt} -SFR \frac{M_{\rm dust, j}}{M_{\rm gas}} - \frac{dM^{\rm SN}_{\rm destr, j}}{dt} -\frac{dM^{\rm YSs}_{\rm destr, j}}{dt}- \\ -\frac{dM^{\rm PN}_{\rm  destr, j}}{dt}-ML \times SFR \frac{M_{\rm dust, j}}{M_{\rm gas}}+\frac{dM_{\rm growth, j}}{dt}.
\end{multlined}
\end{equation}

The first term of the equation represents the dust enrichment where the dust yields are approximated by making use of the metal yields: 
\begin{equation}
\frac{dM^{\rm SP}_{\rm dust, ej, j}}{dt}= \frac{f_{\rm key, j}}{n_{\rm key, j}}\frac{dM^{\rm SP}_{\rm gas, ej, key, j}}{dt} \frac{m_{\rm dust,j}}{m_{\rm key, j}},
\end{equation}

where $f_{\rm key, j}$ is the fraction of key element\footnote{The key element is defined as the least abundant among the elements that form a certain dust species divided by its number of atoms in the compound.} locked into dust grains, $n_{\rm key, j}$ is the number of atoms of the key element in one monomer of dust, and $dM^{\rm SP}_{\rm gas, ej, key, j}/dt$ is the gas mass injection rate of the key element from the SP. We assume $f_{\rm key, olivine}=0.3$, $f_{\rm key, pyroxene}=0.3$, $f_{\rm key, carbon}=0.5$, $f_{\rm key, iron}=0.01$ for TP-AGB stars, and $f_{\rm key, olivine}=0$, $f_{\rm key, pyroxene}=0.5$, $f_{\rm key, carbon}=0.5$, $f_{\rm key, iron}=0.5$, for SNe \citep{Nanni20}. The quantity $m_{\rm key, j}$ is the atomic mass of the key element.
The second term in Eq.~\ref{eq:dust} represents dust astration due to star formation.
The terms $dM^{\rm SN}_{\rm destr, j}/dt$, is the dust destruction from SN shocks. We add the terms $dM^{\rm YSs}_{\rm destr, j}/dt$ and $dM^{\rm PN}_{\rm destr, j}/dt$, corresponding to photo-evaporation due to young stars and PNe, respectively. 
The destruction term in PNe includes both the dust destroyed in the ISM ($dM^{\rm PN, ISM}_{\rm  destr, j}/dt$) and in-situ ($dM^{\rm PN, in-situ}_{\rm  destr, j}/dt$):
\begin{equation}\label{eq:destr_PN}
    \frac{dM^{\rm PN}_{\rm  destr, j}}{dt}=\frac{dM^{\rm PN, ISM}_{\rm  destr, j}}{dt}+\frac{dM^{\rm PN, in-situ}_{\rm  destr, j}}{dt}
\end{equation}
Similarly to Eqs.~\ref{Mgas} and \ref{MZ} dust removal from the outflow is parameterised through the mass-loading factor. The last term is dust growth occurring in the ISM. 

We adopt the commonly-used delayed star-formation history (SFH) for the galaxy:
\begin{equation}\label{SFH}
     SFR \propto \frac{1}{\tau^2}e^{-t/\tau},
\end{equation}
where $\tau=10, 1000$~Myrs are representative of a rapid burst of star formation and of a more continuous star formation, respectively. The SFH is normalised in such a way that  $M_\star=1$~M$_\odot$ after 13~Gyrs. We however checked that the same kind of normalisation at different final ages of the galaxy does not change the results.

\subsection{Dust destruction from SN shocks}
The dust destruction operated by SN shocks is modelled as in many works in the literature, e.g. \citet{Dwek98}. The destruction time-scale is given by:

\begin{equation}\label{Eq:tau_SN}
     \tau_{\rm destr, SN}= \frac{M_{\rm gas}}{R_{\rm SN}(t)M_{\rm swept}},
\end{equation}
where $M_{\rm gas}$ evolves according to Eq.~\ref{Mgas}, $R_{\rm SN}$ is the SN rate and $M_{\rm swept}$ is the mass of gas swept-up for each SN event.
We here assume $M_{\rm swept}=1200$~M$_\odot$ \citep{Dwek07}. 
The destruction rate of dust is therefore:
\begin{equation}\label{Eq:destr_SN}
     \frac{d M^{\rm SN}_{\rm destr, j}}{dt}=\frac{M_{\rm dust, j}}{\tau_{\rm destr, SN}}.
\end{equation}

\subsection{Dust photo-evaporation}\label{Sect:photo_evap}
The stellar parameters of individual stars have been calculated with the FUNS code \citep{Straniero06, Cristallo09, Cristallo11, Cristallo15}. Surface luminosities and temperatures have been extracted when 10\% of the central hydrogen has been burnt.
We consider a metallicity of Z=0.0001 plus enrichment of $\alpha$-elements (0.7 dex, for oxygen, and 0.4 dex for the other $\alpha$-elements) and stellar masses between $0.8$ and $120$~M$_\odot$. However, we verify that by increasing the upper limit of stellar mass up to $300$~M$_\odot$ does not significantly change the results. The low metallicity and stellar evolutionary phase allow for the maximum possible effective temperatures, and therefore the largest possible dust photo-evaporation.
At each distance from the star we compute the dust equilibrium temperature $T_{\rm j}$ for each dust species $\rm j$, either carbon or silicate:

\begin{equation}\label{Eq:eq_temp}
     \int_\nu \kappa_{\rm abs, j}(\nu) B(\nu)(T_{\rm eff}) W(r) d\nu=\int_{\nu}\kappa_{\rm abs, j}(\nu) B(\nu)(T_{\rm j}), d\nu,
\end{equation}
where $\kappa_{\rm abs, j}(\nu)$ is the mass absorption coefficients of the dust species $j$, in cm$^2$ g$^{-1}$ as a function of the frequency, $B(\nu)(T_{\rm eff})$ and $B(\nu)(T_{\rm j})$ are the black body emission at the effective temperature of the star and of the dust temperature, respectively, and $W(r)$ is the dilution factor of radiation with the distance from the star, $r$, which for stars of radius $R_*$ reads:
\begin{equation}
     W(r) = \frac{1}{2}\Bigg[1 - \sqrt{1-\Bigg(\frac{R_*}{r}\Bigg)^2}\Bigg].
\end{equation}

Stars are seldomly born isolated. Therefore, we consider the more realistic case of stars situated in stellar clusters. We assume for the radiation field the one obtained from a simple stellar population (SSP)
computed for a single burst of star formation with $Z=0.0001$ and both the Chabrier and top-heavy IMF by means of the code \textsc{P{\'E}GASE.3} \citep{Fioc19} with a lower and upper limit of 0.8 and 120~$M_\odot$, respectively. We adopt the stellar population  based on ``Padova'' tracks
\citep{Bressan93, Fagotto94a, Fagotto94b, Fagotto94c, Girardi96}.
We consider a zero-age SSP that provides the maximum of the photo-evaporation efficiency, and an upper limit for the star cluster mass of $10^5$~M$_\odot$ \citep{Swart10}.
In this case, Eq.~\ref{Eq:eq_temp} becomes:
\begin{equation}\label{Eq:eq_temp_cluster}
     \int_\nu \kappa_{\rm abs, j}(\nu) \frac{L(\nu)_{\rm SSP}}{4\pi r^2} d\nu=\int_{\nu}\kappa_{\rm abs, j}(\nu) B(\nu)(T_{\rm j}), d\nu,
\end{equation}

where $L(\nu)_{\rm SSP}$ is the spectrum of the SSP.

The mass absorption coefficients in Eqs~\ref{Eq:eq_temp} and \ref{Eq:eq_temp_cluster} for the main populations of silicate and carbon are computed by starting from their optical properties of \citet{Laor93} for astronomical silicate and graphite in the wavelength range from 0.001--1000~$\mu$m which covers the entire emission spectrum of young stars and PNe\footnote{Absorption and scattering coefficients are available at \url{https://www.astro.princeton.edu/~draine/dust/dust.diel.html}}. We adopt the same size distribution of \citet{WD01, DL07}. We do not consider PAHs evolution in our treatment since they usually represent a minor component in terms of the total dust mass in the ISM. For this reason, we exclude from the grain size distribution of carbon the contribution of small grains of size given by Eq.~2 of \citet{WD01}.

We compute the variation of the dust size with time due to evaporation of dust heated by starlight $da/dt^{\rm sub}_{\rm i}$ in analogy to Eq.~3 of \citet{Kobayashi11} and adopting the data in their Table~1:
\begin{equation}
  \frac{da}{dt}^{\rm sub}_{\rm j}=-\frac{1}{\rho_{\rm j}}\sqrt{\frac{A_{\rm j} m_H}{2 \pi k T_{\rm j}}} P_0 \exp\Bigg(-\frac{A_{\rm j} m_H L_{\rm j}}{k T_{\rm j}}\Bigg),
\end{equation}
where $\rho_{\rm j}$ and $\rm L_{\rm j}$ are the mass density and the latent heat
of sublimation of the j-th dust species, $A_{\rm j}$ is its mean molecular weight, $P_0$ is the saturated vapour pressure, $k$ is the Boltzmann constant, and $m_{\rm H}$ is the hydrogen mass. We conservatively adopt the same data of olivine also to compute the evaporation of pyroxene. Indeed, olivine is predicted to evaporate at temperature lower than the one of pyroxene. As a consequence, our choice maximises the evaporation efficiency of silicates.

Small grains evaporate quicker than large ones, therefore, in order to find an upper limit of dust photo-evaporation, we assume that carbon and silicate grains formed have a typical size of $a_{\rm j} = 0.01$~$\mu$m which is about one order of magnitude smaller than the peak of distribution found by \citet{WD01} for the Milky Way and adopted in \citet{DL07}.

We define the dust temperature for which the grains are stable against photo-evaporation as the temperature at which the time needed for evaporation for a grain of size $a_{\rm j} = 0.01$~$\mu$m is greater or equal to a critical age, $\rm age_{\rm crit}$, which we set equal to the age at which the galaxy has built its entire stellar mass, $\rm age_{\rm crit}= 13$~Gyrs:
\begin{equation}
    \frac{a_{\rm j}}{da/dt^{\rm sub}_{\rm j}}={\rm age}_{\rm crit} 
\end{equation}
We obtain $T_{\rm carbon}\approx 1330$~K for carbon dust and $\rm T_{\rm silicate}\approx 877$~K for silicates. 

The distance from the source at which the equilibrium temperature equals this sublimation temperature (Eqs.~\ref{Eq:eq_temp} and \ref{Eq:eq_temp_cluster}) is the dust sublimation radius, $R_{\rm subl, j}$.

In case of individual stars (ISs) or PNe, at each time-step the mass of dust destroyed in the volume of a shell of thickness equal to $R^{\rm IS/PN}_{\rm subl, j}-R_*$ is:
\begin{equation}\label{Eq:Mdestri}
      M^{\rm IS/PN}_{\rm destr, j} = \frac{M_{\rm dust, j}}{M_{\rm gas}} \frac{4}{3}\pi \Big(R^3 _{\rm subl, j}-R_*^3\Big) \rho_{\rm gas}.
\end{equation}
For stars in the ISM, $\rho_{\rm gas}$ is either the gas density in the region of star formation or where the PN resides. For young stars, we set as density that of a molecular cloud $\rho_{\rm gas}=10^5 m_{\rm H}\mu$ where $\mu$ is the mean molecular weight $\mu=1.37$. Such a high value of the density is typical of star-forming clumps and cores \citep{Draine11}. The contribution of the destruction by each isolated star is weighted over the IMF (which is normalised as in Eq.~\ref{eq:imf_norm}):
\begin{equation}\label{Eq:Mdestri_2}
      M^{\rm IS, W}_{\rm destr, j} = \int^{M_{\rm U}}_{M_{\rm L}}IMF(m) \frac{M_{\rm dust, j}}{M_{\rm gas}} \frac{4}{3}\pi \Big(R^3 _{\rm subl, j}-R_*^3\Big) \rho_{\rm gas} dm.
\end{equation}

For the calculation of the dust destroyed by a single PN in the ISM  we use Eq.~\ref{Eq:Mdestri} and assume $\rho_{\rm gas}=50 m_{\rm H}\mu$, a value typical of the cold atomic medium \citep{Ferriere01} which is an upper limit of the density of the medium in which PNe are expected to reside. The quantity $M_{\rm dust, j}/M_{\rm gas}$ is the dust-to-gas ratio in the ISM calculated from Eqs.~\ref{Mgas} and \ref{eq:dust}.

Planetary Nebulae however are complex objects in which the central hot white dwarf is surrounded by dust and gas at higher density than the one of the cold atomic medium. Dust is produced during the TP-AGB phase when the star loses mass at high rates (up to a few $ 10^{-5}$~$M_\odot yr^{-1}$). 
For computing dust destruction in-situ for PNe we assume $\rho_{\rm gas}=10^4 m_{\rm H}\mu$ \citep[e.g.][]{Stanghellini12} and an upper limit of the total dust-to-gas ratio of 0.01 \citep{Stasinska99}. We assume that each of the dust species considered (olivine, pyroxene and carbon) has $M_{\rm dust, j}/M_{\rm gas}$ equal to one third of the total value.

In case the stars are born in clusters we assume that the radiation comes from a point source. In this case, the mass of dust destroyed is the one in a sphere of radius $R^{\rm SP}_{\rm subl, j}$:
\begin{equation}\label{eq:destr_SSP}
     M^{\rm SP}_{\rm destr, j} = \frac{M_{\rm dust, j}}{M_{\rm gas}} \frac{4}{3}\pi R^3_{\rm subl, j} \rho_{\rm gas},
\end{equation}

where $\rho_{\rm gas}$ is assumed to be the same as in the case of individual stars ($\rho_{\rm gas}=10^5 m_{\rm H}\mu$).

The final dust destruction operated by star formation is therefore computed given the SFR as a function of time:
\begin{equation}\label{Eq:SF_destr}
 \frac{d M^{\rm YSs}_{\rm destr, j}}{dt}= \frac{M^{\rm IS, W/SP}_{\rm destr, j}}{M_{\rm gas}} SFR,
\end{equation}

Photo-evaporation rate induced by PNe is instead calculated as: 
\begin{equation}\label{Eq:PN_destr}
     \frac{d M^{\rm PN, ISM/in-situ}_{\rm destr, j}}{dt} =  M^{\rm PN, ISM/in-situ}_{\rm destr, j} \frac{dn_{\rm AGB}}{dt},
\end{equation}

where $dn_{\rm AGB}/dt$ is the AGB ``birth'' rate that end their evolution as PNe which is extracted from the code \textsc{omega}. We assume an effective temperature for all the PNe of $T_{\rm eff}=2\times 10^5$~K and luminosity $L=4 \times 10^4$~L$_\odot$ which are upper limits values for PNe with initial mass of 6~M$_\odot$ \citep[see FRUITY database\footnote{\url{fruity.oa-abruzzo.inaf.it}},][]{Cristallo15}.

\subsection{Dust growth in the ISM}
The dust destruction processes included in our calculations depends on the dust-to-gas ratio as a function of time. Since such a quantity varies if dust growth in the ISM is included, we analyse two extreme cases: 1) no dust growth in the ISM; 2) fully efficient dust growth in the ISM.
The variation of mass of dust due to growth is expressed as:

\begin{equation}
     \frac{dM_{\rm growth, j}}{dt}=4\pi\frac{da_{\rm j}}{dt}a_{\rm j}^2 \rho_{\rm j} n_{\rm s, j},
\end{equation}
where $\rm da_{\rm j}/dt$ is the variation of the dust size due to the addition of atoms and molecules on the grain surface and is computed following \citet{Nanni20}, $n_{\rm s, j}$ is the number of seed nuclei computed by dividing the mass of each dust species by the mass of one dust grain \citep{Asano13}. For this calculation we implicitly assume that all the grains in the ISM can potentially act as seed nuclei.

\section{Results}\label{results}
\subsection{A simple case}
In order to give a sense of how much dust is destroyed for each solar mass of stars formed and for each PN, we here perform some simple calculations by considering a single value for the dust-to-gas ratio of $0.01$ in the ISM \citep{Bohlin78}. As sublimation radius of dust in Eqs.~\ref{Eq:Mdestri}, \ref{Eq:Mdestri_2} and \ref{eq:destr_SSP} we adopt the one of silicates, $R_{\rm subl, silicate}$, since this is larger than the one of carbonaceous grains. 
\subsubsection{Dust photo-evaporation by isolated young stars}\label{single_stars}
Based on Eq.~\ref{Eq:Mdestri_2} we first estimate how much dust is destroyed per $M_{\odot}$ of stars formed with the IMF distributed according to the Chabrier or top-heavy functions and normalised as in Eq.~\ref{eq:imf_norm}. For this test, we assumed stars to be formed in isolation. We obtain the fraction of $\approx 1.2\times 10^{-8}$ and $\approx 6.7\times 10^{-8}$ of dust destroyed, where the stars are distributed according to the Chabrier and top-heavy IMF, respectively.

\subsubsection{Dust photo-evaporation by stars in clusters}\label{clusers}
From Eq.~\ref{eq:destr_SSP} we find that the dust mass fraction destroyed per solar mass of stars formed in a cluster is $\approx 1.0\times 10^{-8}$ and $\approx 1.3\times 10^{-7}$ where stars are distributed according to the Chabrier and top-heavy IMF, respectively. 
In the case of the top-heavy IMF, such an estimate is larger than the one found for isolated stars, while the two values are comparable for the Chabrier IMF. The difference with the isolated stars case may be due to the diverse estimates of flux impinging on the grain surface in Eqs.~\ref{Eq:eq_temp} and \ref{Eq:eq_temp_cluster}, and of the total volume cleared-up from dust in Eqs.~\ref{Eq:Mdestri} and \ref{eq:destr_SSP}.
 
The sublimation radius for the different dust species are $R_{\rm subl, silicate}\approx 9\times10^{-3}$~pc, and $R_{\rm subl, carbon}\approx 5\times10^{-3}$~pc.
These values suggest that dust is destroyed by photo-evaporation in the immediate vicinity of the cluster, given that a typical radius of an HII region is of a few pc \citep{Draine11}.

\subsubsection{Dust photo-evaporation by PNe}\label{PN}
We estimate from Eq.~\ref{Eq:Mdestri} that the amount of dust destroyed in the ISM for each PN is $\approx2.4\times 10^{-11}$~M$_\odot$.

We obtain that the mass of dust in-situ destroyed by photo-evaporation is $\approx 4.9\times 10^{-9}$~$M_\odot$. Such a value is negligible compared to the dust yields produced along the entire duration of the TP-AGB phase \citep[e.g.][]{Zhukovska08, Ventura12,Nanni13}, and to the dust mass inferred from observations which is of the order of $\approx 10^{-4}$~$M_\odot$ \citep{Stasinska99, Dellagli23}. Recently, \citet{Dellagli23} have found that the amount of dust possibly destroyed by PN feedback is up to $\approx 60\%$. Therefore, an alternative mechanism different from photo-evaporation may be at work. As for the case of clusters, photo-evaporation is efficient in the vicinity of the source. We find $R_{\rm subl, silicate}\approx 7\times10^{-4}$~pc, and $R_{\rm subl, carbon}\approx 9\times10^{-5}$~pc.

\subsection{Dust exposed to an interstellar radiation field- (ISRF-) like radiation}
Dust in the ISM as well as in the outer part of the cirmcumstellar envelopes of TP-AGB stars, is exposed to a background radiation field.
We here assess if photo-evaporation may be relevant when dust is exposed to an external radiation field as often assumed in the literature \citep[e.g.][]{DL07}. In such an approach, the radiation field is parameterised as $U\times ISRF$ where $U$ is a scaling which can be up to $U_{\rm max}=10^7$ and the ISRF is from \citet{Mathis83}. For such a maximum value of the scaling factor we obtain from Eq.~\ref{Eq:eq_temp} \cite[by substituting $B(\nu)(T_{\rm eff})$ with the ISRF flux by][]{Mathis83} an equilibrium temperature for silicate dust of $T_{\rm silicate}\approx154$~K and of $T_{\rm carbon}\approx 245$~K for carbon. Such values are well below the sublimation temperatures derived in Section~\ref{Sect:photo_evap} and therefore both silicate and carbon dust grains are stable against evaporation.

\subsection{Including dust destruction in dust evolutionary models}
\begin{figure}
   \centering
   \includegraphics[scale=0.35]{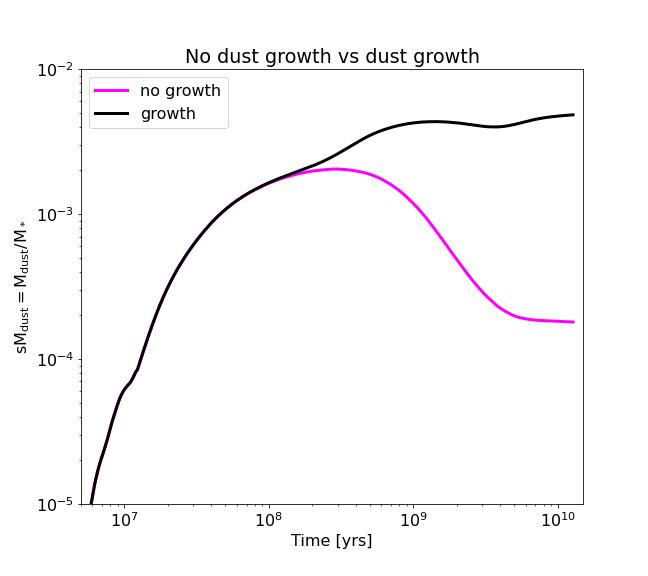}
   \includegraphics[scale=0.35]{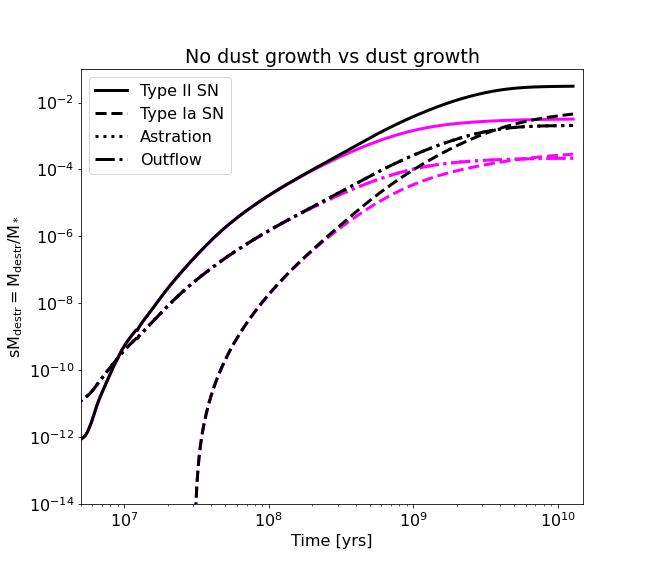}
   \includegraphics[scale=0.35]{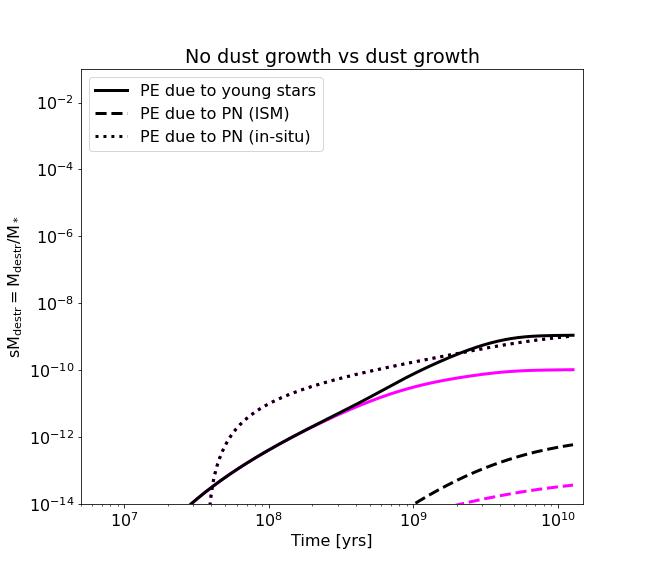}
   \caption{Upper panel: overall evolution of dust computed as the enrichment from stars minus the dust removal from different mechanisms. The cases with and without dust growth in the ISM are shown. The initial mass of the gas is equal to $4$ times the final mass of stars, $M_{\rm gas, ini}= 4\times M_{\rm \star , fin}$ and $\tau=1000$~Myrs in Eq.~\ref{SFH}. Middle and lower panels: dust removal due to different mechanisms included in the models in the upper panel as mentioned in the legend, where ``PE'' stands for ``photo-evaporation''. }
              \label{Fig:destr_tau}
   \end{figure}

We include dust photo-evaporation by considering the case of young stars born in clusters and PNe, following the description provided in Section~\ref{method}.
In Fig.~\ref{Fig:destr_tau} we show the overall evolution of $sM_{\rm dust}$ in the ISM (upper panel) and we compare the efficiency of different dust depleting mechanisms by plotting the integrated amount of the specific mass of dust destroyed or removed as a function of time, $sM_{\rm destr}=M_{\rm destr}/M_{*}$ (middle and lower panels). In other words, we integrate in time each of the destruction mechanisms corresponding to the terms from the 2nd to the 6th in Eq.~\ref{eq:dust} and we divide by the stellar mass\footnote{The contribution of photo-evaporation from PNe has been splitted in the two terms of Eq.~\ref{eq:destr_PN}.}. We run a reference model by assuming $M_{\rm gas, ini}=4$~$M_{\rm *, fin}$ (where $M_{\rm gas, ini}$ and M$_{\rm *, fin}$ are the initial mass of gas and the final stellar mass, respectively), $\tau=1000$~Myrs in Eq.~\ref{SFH}, $M_{\rm swept}=1200$~M$_\odot$, a Chabrier IMF, and a typical value of the mass-loading factor $ML=1$. Note that with such an assumption for $ML$ astration and outflow provide the same contribution to dust removal. We consider the cases with and without dust growth in the ISM, and we vary $M_{\rm gas, ini}$, $\tau$ and the IMF. 

Fig.~\ref{Fig:destr_tau} demonstrates that destruction from photo-evaporation is orders of magnitudes lower than SN shocks destruction, than dust astration, and than outflow removal. This is true both with and without dust growth in the ISM.
The difference in the destruction efficiency for the two cases is minor. Only the model in which dust growth is not included shows a decline in $sM_{\rm dust}$ as a function of time mainly due to dust destruction from Type II SNe.
Such a trend is in qualitative good agreement with the one observed in the literature, however, a number of observable parameters like metallicity and gas fraction, beside specific mass of dust should be reproduced consistently by models before drawing some firm conclusion.

\begin{figure}
   \centering
   \includegraphics[scale=0.35]{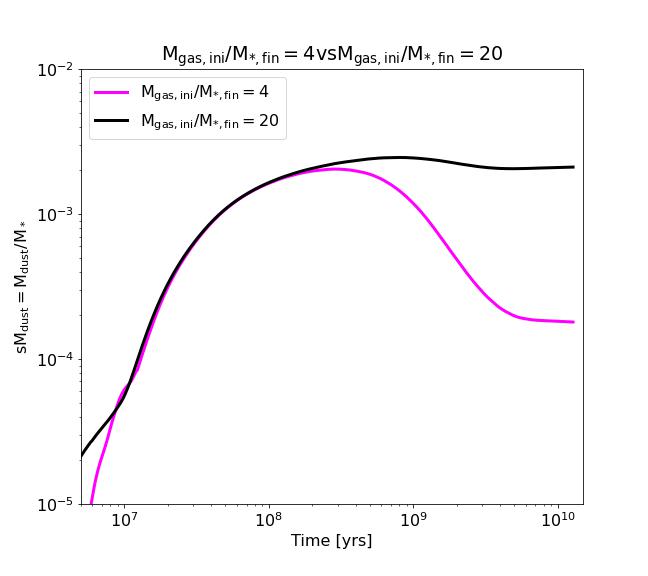}
   \includegraphics[scale=0.35]{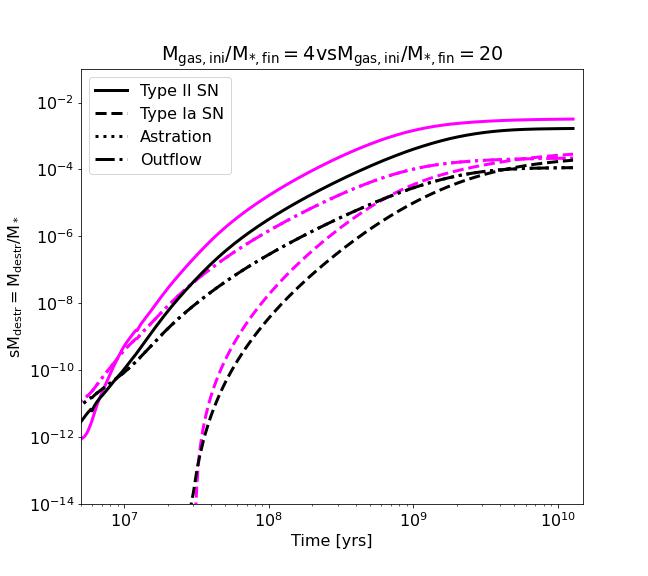}
   \includegraphics[scale=0.35]{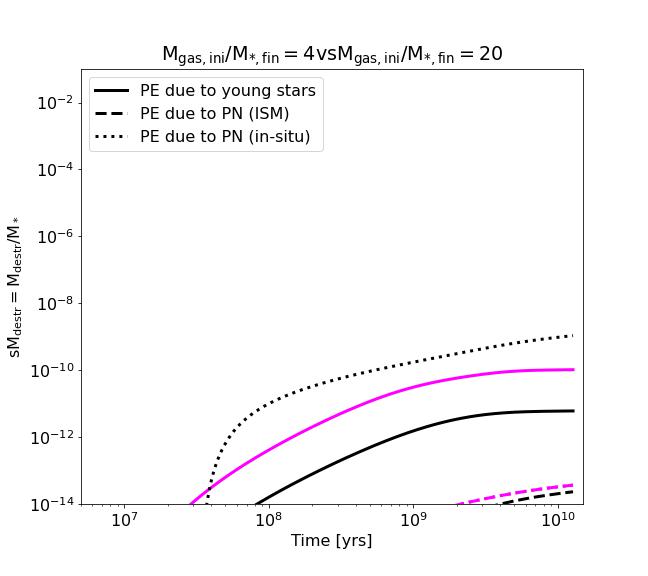}
      \caption{Upper panel: overall evolution of dust computed as the enrichment from stars minus the dust removal from different mechanisms. Two cases with different initial gas masses, $M_{\rm gas, ini}= 4\times M_{\rm \star, fin}$ and $M_{\rm gas, ini}= 20\times M_{\rm \star, fin}$, are compared. We select $\tau=1000$~Myrs in Eq.~\ref{SFH} and we do not include dust growth in the ISM. Middle and lower panels: dust removal due to different mechanisms included in the models in the upper panel, similar to Fig.~\ref{Fig:destr_tau}. }
              \label{Fig:destr_mg10}
   \end{figure}

In Fig.~\ref{Fig:destr_mg10} we show the same model as in Fig.~\ref{Fig:destr_tau} (magenta lines) together with a model representative of gas rich galaxies with $M_{\rm gas, ini}= 20\times M_{\rm *, fin}$. In the right panel we can appreciate that in this latter case the efficiency of the various removal processes is reduced due to the effect of lower values of $M_{\rm dust, j}/M_{\rm gas}$ (see Eqs.~\ref{Eq:tau_SN}, \ref{Eq:destr_SN} and \ref{Eq:Mdestri}). Photo-evaporation remains negligible with respect to the other processes.
The plot shows that for gas rich systems with large $M_{\rm gas}/M_{\star}$ the predicted $sM_{\rm dust}$ does not show a decline as instead in the case with a smaller $M_{\rm gas}/M_{\star}$.

\begin{figure}
   \centering
   \includegraphics[scale=0.4]{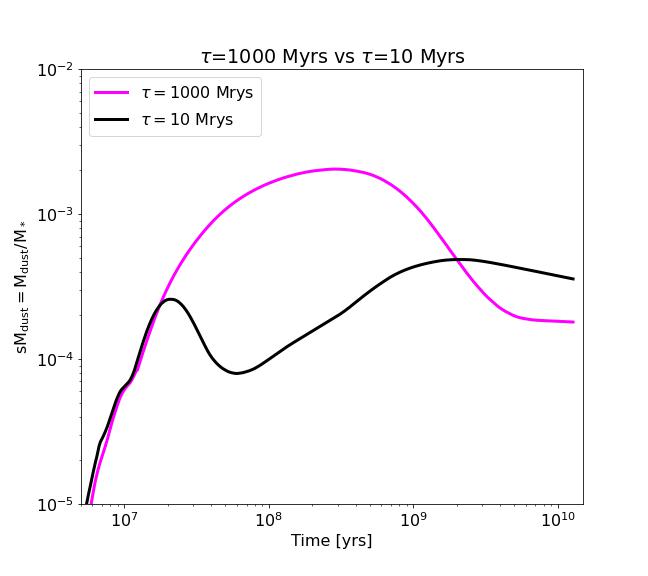}
   \includegraphics[scale=0.4]{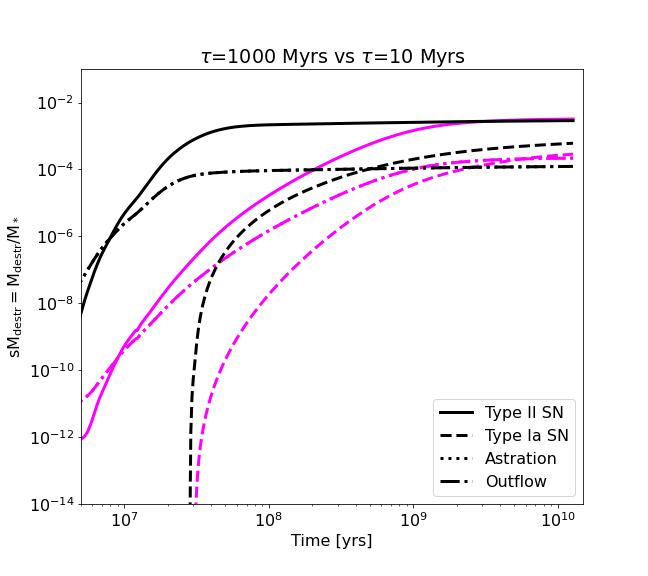}
   \includegraphics[scale=0.4]{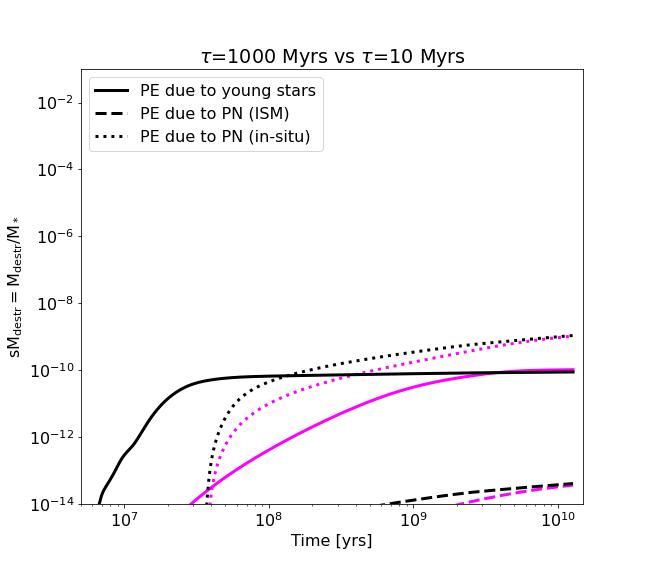}

   \caption{Upper panel: overall evolution of dust computed as the enrichment from stars minus the dust removal from different mechanisms. Two cases with different values of $\tau=10,1000$~Myrs in Eq.~\ref{SFH} are shown. $M_{\rm gas, ini}= 4\times M_{\rm \star, fin}$. Middle and lower panels: dust removal  due to different mechanisms corresponding to the models in the upper panel, similar to Fig.~\ref{Fig:destr_tau}. }
              \label{Fig:destr_tau10}
   \end{figure}

In Fig.~\ref{Fig:destr_tau10} we show the same model as in the previous plots (magenta lines) together with a model computed by assuming $\tau=10$~Myrs in Eq.~\ref{SFH} representing the case of a very short burst of star formation. In this latter case, the decline in the $sM_{\rm dust}$ with time occurs at earlier times with respect to the case with $\tau=1000$~Myrs due to the fast production of stars which explodes as Type II SN, and it is followed by an increase in the dust content due to the remaining contribution of Type II SNe, AGB stars and Type Ia SNe. Such a trend showing a decline followed by an increase in the dust mass seems to be at odd with the observed trend, but is obtained for the specific choice of parameters (e.g $M_{\rm gas, ini}/M_{\rm \star, fin}$) beside the SFH.

\begin{figure}
   \centering
   \includegraphics[scale=0.4]{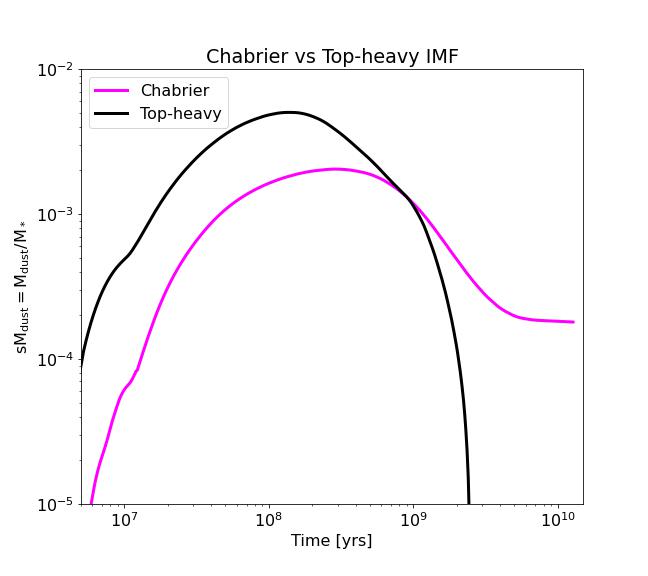}
   \includegraphics[scale=0.4]{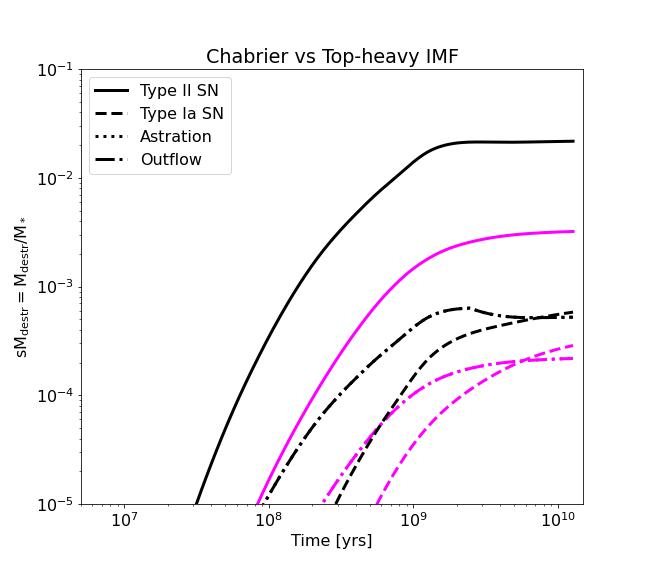}
   \includegraphics[scale=0.4]{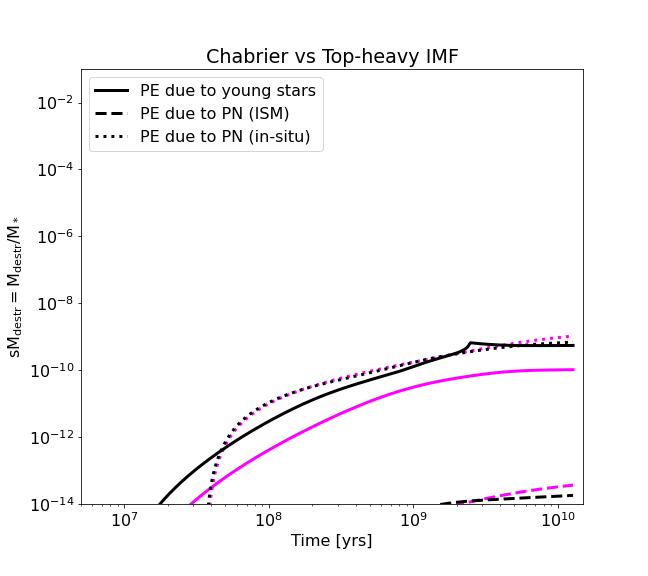}

   \caption{Upper panel: overall evolution of dust computed as the enrichment from stars minus the dust removal from different mechanisms. Two cases with different choices of the IMF are shown. We select $M_{\rm gas, ini}= 4\times M_{\rm \star, fin}$, $\tau=1000$~Myrs in Eq.~\ref{SFH} and we do not include dust growth in the ISM. Middle and lower panels: dust removal due to different mechanisms corresponding to the models in the upper panel, similar to Fig.~\ref{Fig:destr_tau}.  
   }
              \label{Fig:destr_IMF}
   \end{figure}

In Fig.~\ref{Fig:destr_IMF} we show the effect of assuming different IMF (either \citet{Chabrier03} or top-heavy). A top-heavy IMF may be typical of low-metallicity environments \citep[e.g.][]{Marks12}. In the upper panel, the model with the top-heavy IMF shows a more rapid decline with time with respect to the Chabrier case due to the efficient destruction by Type II SNe which are more numerous than for the Chabrier IMF. The dust produced by the top-heavy model, encompasses the one in which the Chabrier IMF is employed. Because of the larger $sM_{\rm dust}$ the destruction mechanisms are more efficient than in the Chabrier case. Photo-evaporation remains negligible in both cases. For both the models, the trend found is in qualitative good agreement with the observations.

\section{Conclusions}
We find that photo-evaporation due to young stars and PNe has only a minor role in dust destruction, independently of the assumed efficiency of dust growth in the ISM, the initial mass of gas, the SFH, and the IMF. We do not exclude however the possibility of dust being destroyed by shocks generated by fast winds in HII regions and PNe. The investigation of such processes is beyond the scope of this work.
The dust exposed to an interstellar radiation field $U\times ISRF$ is stable against sublimation up to the largest value assumed in the literature $U_{\rm max}=10^7$ \citep{DL07}. These findings also imply that the total yields from TP-AGB stars are not normally destroyed by photo-evaporation induced by the ambient radiation field or when the PN phase is reached.

The observed trend with increasing $sM_{\rm dust}$ at early times followed by a decrease is qualitatively well reproduced for low values of $M_{\rm gas}/M_{\star}$ found in passive galaxies, while for larger values of $ sM_{\rm dust}$ representative of systems with a large gas fraction (e.g. DGS) the observed trend is difficult to be reproduced if outflow is not extremely efficient (see also \citet{Nanni20}). 

It is worth noticing that observationally we are able to probe specific dust mass above $10^{-4}$ at each redshift. Therefore, at such values we don't expect photo-evaporation to be a relevant dust destruction process even with low metallicity and intense radiation field.

\bibliographystyle{aa}
\bibliography{nanni.bib}
\begin{acknowledgements}
A.N, M.R., P.S. acknowledge support from the Narodowe Centrum Nauki (NCN), Poland, through the SONATA BIS grant UMO-2020/38/E/ST9/00077. 
D.D. acknowledges support from the NCN through the SONATA grant UMO-2020/39/D/ST9/00720.
M.J.M.~acknowledges the support of 
the NCN through the SONATA BIS grant UMO-2018/30/E/ST9/00208 and the Polish National Agency for Academic Exchange Bekker program grant BPN/BEK/2022/1/00110. 
M.R. acknowledges support from the Foundation for Polish Science (FNP) under the program START 063.2023.
We thank the anonymous referee for the careful reading of the manuscript and for her/his thoughtful comments.
\end{acknowledgements}
\end{document}